\documentclass[twocolumn,pra,aps,showpacs]{revtex4}

\usepackage{mathptmx}
\usepackage{subfigure}
\usepackage{psfrag,graphicx}
\usepackage{dcolumn}
\usepackage{amsmath,amssymb}
\usepackage{bm}
\usepackage{color}
\usepackage{latexsym}
\usepackage{epstopdf}
\usepackage{color}
\usepackage[english]{babel}
\usepackage{latexsym}
\usepackage{psfrag,graphicx}
\usepackage{subfigure}
\usepackage{amsmath}
\usepackage{amssymb}
\usepackage{amsfonts}
\usepackage{bm}
\usepackage{natbib}
\usepackage{epstopdf}
\usepackage{appendix}

\definecolor{mygrey}{gray}{0.35}
\definecolor{myblue}{rgb}{0.2,0.2,0.8}
\definecolor{myzard}{cmyk}{0,0,0.05,0}
\definecolor{mywhite}{rgb}{1,1,1}
\definecolor{mywhite}{rgb}{1,1,1}
\definecolor{myred}{rgb}{1,0.,0.3}

\usepackage[colorlinks=true,citecolor=myblue,linkcolor=myred]{hyperref}

\def\ba{\begin{align}}
\def\enda{\end{align}}
\def\bi{\begin{itemize}}
\def\ei{\end{itemize}}

\def\be{\begin{equation}}
\def\ee{\end{equation}}
\def\bea{\begin{eqnarray}}
\def\eea{\end{eqnarray}}
\def\bse{\begin{subequations}}
\def\ese{\end{subequations}}

\begin{document}
\title{Quantum Thermometry with Trapped Ions}
\author{Peter A. Ivanov}
\affiliation{Department of Physics, St. Kliment Ohridski University of Sofia, James Bourchier 5 blvd, 1164 Sofia, Bulgaria}

\begin{abstract}
We introduce the estimation protocol for detecting the temperature of the transverse vibrational modes of linear ion crystal. We show that thanks to the laser induced laser coupling between the vibrational modes and the collective spin states the estimation of the temperature is carried out by set of measurements of the spin populations. We show that temperature estimation protocol using single ion as a quantum probe is optimal in a sense that the set of state projective measurement saturates the fundamental Cramer-Rao bound. We find a plateau of the maximal temperature sensitivity using ion chain as a quantum probe. Moreover, we show that the non-classical part of the quantum Fisher information could leads to enhancement of the temperature sensitivity compared to the single ion case.
\end{abstract}

\maketitle

\section{Introduction}
Precise temperature estimation has attracted recently considerable interest due to the broad range of technological applications including medicine and biology \cite{Klinkert2009} as well as quantum thermodynamics \cite{Gemmer2004}. For example measuring low temperature with high precision in a controlled quantum systems can be used to test and explore the thermodynamics in the quantum regime \cite{An2014,Rosnagel2016}. One way to determine the temperature of the quantum system is to measure its energy where the ultimate limit of estimation precision was recently discussed \cite{Stace2010,Paris2016,Correa2015}. Other approach is based on the coupling between the system at the thermal equilibrium and ancillary probe qubit system where the temperature estimation is carried out by state-dependent measurement of the qubit states \cite{Brunelli2011,Brunelli2012,Higgins2012}. Various quantum systems can be used to detect very low temperature including for example Bose-Einstein condensate \cite{Sabin2014} and ultracold lattice gases \cite{Mehboundi2015}. Another promising quantum system with application in low temperature measurement is the laser cooled trapped ions which provide excellent experimental control over the external and internal degree of freedom as well as high fidelity read out of the quantum state \cite{Turchette2000,Singer2015}.

In this work we consider the estimation on the temperature of the two transverse vibrational modes of linear ion crystal via state dependent measurement of the collective ion's spin states. We show that the bichromatic laser field can be used to couple the transverse vibrational modes with the collective spin states with tunable coupling strength and detuning. Our temperature estimation protocol consists of Ramsey type sequence where the temperature dependent phase acquired by the spins during the time evolution is mapped on the respective spin populations which are subsequently measured. We quantify the estimation precision in terms of classical and quantum Fisher information. For the single ion case we show that the estimation protocol is optimal in a sense that it leads to equality between the classical and quantum Fisher information. We find the optimal spin basis and show that is independent on the temperature. We extend the discussion by considering multi-ion chain as a quantum probe. At low temperature limit we find a plateau of the maximal sensitivity compared to the single ion case. We examine the quantum Fisher information and show that its non-classical part could leads to enhancement of the temperature sensitivity.

The paper is organized as follows: In Sec. \ref{q_probe} we discuss the vibrational modes of the linear ion crystal. In Sec. \ref{LII} we consider the laser-ion interaction which coupled the transverse vibrational modes with the collective spin states. In Sec. \ref{PHE} the adiabatic elimination of the phonon states is discussed. We show that depending on the sign of the laser detuning the time evolution generated by the residual spin-phonon interaction takes the form of spin-dependent beam-splitter (BS) operator or two-mode squeezing (TMS) operator. In Sec. \ref{TEGB} we provide the general background of the temperature estimation. In Sections \ref{TEBSO} and \ref{TETMSO} we consider the Ramsey interferometry sequence in which the information of the temperature is mapped on the spin populations. Finally, in Sec. \ref{DO} we summarize our findings.
\section{Quantum Probe}\label{q_probe}
We begin by considering ion system which consists of $N$ ions with charge $e$ and mass $m$ confined in linear Paul trap with trapped frequencies $\omega_{\alpha}$ ($\alpha=x,y,z$). The potential energy of the ion system is a sum of the effective harmonic potential and the mutual Coulomb interaction given by \cite{Wineland1998,Hafner2008}
\begin{equation}
\hat{V}=\frac{m}{2}\sum_{\alpha=x,y,z}\sum_{j=1}^{N}\omega_{\alpha}^{2}\hat{r}_{\alpha,j}^{2}
+\sum_{i>j}^{N}\frac{e^{2}}{|\hat{\vec{r}}_{i}-\hat{\vec{r}}_{j}|}.
\end{equation}
For sufficiently strong transverse confinement with trap frequencies $\omega_{x(y)}\gg\omega_{z}$ the ions are arranged in a linear configuration along the trap axis $z$. The equilibrium ion's position $z_{i}^{(0)}$ are determined by the balance between the Coulomb repulsion force and the harmonic trapping force, which are quantified by the condition $(\partial \hat{V}/\partial \vec{r}_{i})_{r_{i}=z_{i}^{(0)}}=0$.
Hereafter we consider the low temperature limit where we can expressed the position operator of ion $j$ as $\hat{\vec{r}}_{j}=(z_{j}^{0}+\delta \hat{r}_{z,j})\vec{e}_{z}+\delta \hat{r}_{x,j}\vec{e}_{x}+\delta \hat{r}_{y,j}\vec{e}_{y}$ where the displacement operator $\delta \hat{r}_{\alpha,j}$ describes the quantum harmonic oscillation of the ion around the equilibrium position. Within the harmonic approximation the vibrational Hamiltonian becomes
\begin{equation}
\hat{H}_{\rm vib}=\sum_{\alpha}\sum_{j=1}^{N}\frac{\hat{p}_{\alpha,j}^{2}}{2m}+\frac{m}{2}\sum_{\alpha}\sum_{j,l=1}^{N}\omega_{\alpha}^{2}K_{j,l}^{(\alpha)}
\delta \hat{r}_{\alpha,j}\delta \hat{r}_{l,j},\label{H_c}
\end{equation}
where $K_{j,l}^{(\alpha)}$ is the spring-constant matrix \cite{James1998,Marquet2003}. Since the Hamiltonian (\ref{H_c}) is quadratic in the momentum and displacement operators one can diagonalize it by introducing collective vibrational normal modes. Indeed, solving the eigenvalue problem, $\sum_{j=1}^{N}K_{p,j}^{(\alpha)}b_{p,n}^{(\alpha)}=\lambda_{\alpha,n}b_{j,n}^{(\alpha)}$, where $b_{p,n}^{(\alpha)}$ is the component of the $n$th normal mode eigenvector and $\lambda_{\alpha,n}$ is the corresponding eigenvalue one can expressed the displacement operators in terms of normal modes as $\delta \hat{r}_{\alpha,i}=\sum_{n=1}^{N}b_{i,n}^{(\alpha)}\sqrt{\hbar/2m\omega_{\alpha,n}}(\hat{a}_{\alpha,n}^{\dag}+\hat{a}_{\alpha,n})$. Here $\omega_{\alpha,n}=\omega_{\alpha}\sqrt{\lambda_{\alpha,n}}$ is the collective vibrational frequency and $\hat{a}_{\alpha,n}^{\dag}$, $\hat{a}_{\alpha,n}$ are the respective creation and annihilation operators of phonon in the $n$th mode and direction $\alpha$. Finally, the vibrational Hamiltonian becomes
\begin{equation}
\hat{H}_{\rm vib}=\hbar\sum_{\alpha}\sum_{n=1}^{N}\omega_{\alpha,n}\left(\hat{a}_{\alpha,n}^{\dag}\hat{a}_{\alpha,n}+\frac{1}{2}\right).
\end{equation}
We emphasize that for $n=1$ the collective frequencies becomes $\omega_{\alpha,1}=\omega_{\alpha}$ with normal mode eigenvectors $b_{i,1}^{(\alpha)}=1/\sqrt{N}$. In this mode all ions oscillate in the same manner which we refer it as collective center-of-mass (c.m.) motion.

In the following we discuss measurement of the temperature of the c.m. modes in two orthogonal transverse $x$-$y$ directions by detecting the population of the internal ion's states. For this goal we consider the laser-ion interaction which coupled the ion's internal electronic states with the c.m. vibration modes. We show for sufficiently high effective phonon frequencies compared to the spin phonon couplings the c.m. vibrational modes can be adiabatically eliminated such that the residual spin-phonon interaction takes the form of the spin-dependent phonon beam-splitter \cite{Campos1989} as well as two mode squeezing operators \cite{Caves1991}. The temperature estimation is performed by mapping the temperature dependent phase acquired by the spin states during the Ramsey sequence into the spin-state populations.

\section{Laser-Ion Interaction}\label{LII}
Consider that each ion has two metastable internal states $\left|\uparrow\right\rangle$, $\left|\downarrow\right\rangle$ with Bohr frequency difference $\omega_{0}$. The interaction-free Hamiltonian describing the internal and external degrees of freedom is given by
\begin{equation}
\hat{H}_{\rm free}=\hbar\omega_{0}\hat{J}_{z}+\hat{H}_{\rm vib},
\end{equation}
where we have introduced collective spin operators $\hat{J}_{\alpha}=\frac{1}{2}\sum_{k=1}^{N}\sigma_{k}^{\alpha}$ with $\sigma_{k}^{\alpha}$ being the Pauli operator for $k$th spin.

In order to couple the internal spin states with the collective vibrational states we assume that bichromatic laser fields are applied along the transverse $x$ and $y$ directions with laser frequencies $\omega_{r,\alpha}=\omega_{0}-(\omega_{x}+\delta_{x})$ and $\omega_{b,\alpha}=\omega_{0}+(\omega_{x}-\delta_{x})$ \cite{Lee2005,Schneider2012}. Here $\delta_{x}$, $\delta_{y}$ with $\delta_{x(y)}\ll\omega_{x(y)}$ are the laser detunings to the c.m. vibrational modes along the two transverse directions. The bichromatic laser field causes simultaneous excitation of the red- and blue-sideband transitions between the spin and motion states. The total Hamiltonian becomes $\hat{H}=\hat{H}_{0}+\hat{H}_{I}$ with
\begin{eqnarray}
\hat{H}_{I}&=&\hbar\sum_{k=1}^{N}\sum_{\alpha=x,y}\Omega_{\alpha}\{\left|\uparrow_{k}\right\rangle\left\langle\downarrow_{k}\right|e^{i k_{\alpha}\delta \hat{r}_{\alpha,k}-i\phi_{\alpha}}(e^{-i\omega_{r,\alpha}t}+e^{-i\omega_{b,\alpha}t})\notag\\
&&+{\rm h.c.}\},
\end{eqnarray}
where $\Omega_{\alpha}$ is the Rabi frequency, $\vec{k}_{\alpha}$ is the laser wave vector, and $\phi_{\alpha}$ is the respective laser phase. We introduce Lamb-Dicke parameter $\eta_{\alpha}=k_{\alpha}\sqrt{\hbar/2m\omega_{\alpha}}$ and assume Lamb-Dicke limit $\eta_{\alpha}\ll 1$. Transforming the total Hamiltonian in an interaction picture with respect to $\hat{U}(t)=e^{-i\omega_{0}t\hat{J}_{z}-i\sum_{\alpha}\sum_{p=1}^{N}(\omega_{\alpha,p}-\delta_{\alpha})t}$ we obtain
\begin{eqnarray}
&&\hat{H}=\hat{H}_{\rm b}+\hat{H}_{\rm sb},\quad \hat{H}_{\rm b}=\hbar\sum_{\alpha=x,y}\delta_{\alpha}\hat{a}_{\alpha}^{\dag}\hat{a}_{\alpha},\notag\\
&&\hat{H}_{\rm sb}=\hbar\sum_{\alpha=x,y}\frac{2g_{\alpha}}{\sqrt{N}}\hat{J}_{\alpha}
(\hat{a}_{\alpha}^{\dag}+\hat{a}_{\alpha}),\label{Hsb}
\end{eqnarray}
where $g_{\alpha}=\eta_{\alpha}\Omega_{\alpha}$ is the spin-phonon coupling. The term $\hat{H}_{\rm b}$ describes a quantum harmonic oscillators with effective frequency $\delta_{\alpha}$. The term $\hat{H}_{\rm sb}$ describes the desired coupling between the collective spin operators and the c.m. vibrational modes. In the expression (\ref{Hsb}) we have assumed that all vibrational modes can be neglected except the c.m. mode, which is justified as long as $(\omega_{\alpha}-\omega_{\alpha,p\neq 1})\gg g,|\delta_{\alpha}|$. For simplicity hereafter we assume equal couplings, $g_{\alpha}=g$.

Finally, we point out because each spin is equally coupled to the c.m. vibrational mode in $x$-$y$ directions one can introduce the collective spin basis spanned by the Dicke states $\left|j,m\right\rangle$, which are simultaneous eigenvectors of $\hat{J}^{2}\left|j,m\right\rangle=j(j+1)\left|j,m\right\rangle$ and $\hat{J}_{z}\left|j,m\right\rangle=m\left|j,m\right\rangle$, where $j=N/2$ is the length of the maximum spin of the system. Including the motion degree of freedom the total Hilbert space is spanned by the vectors $|j,m\rangle|n_{x},n_{y}\rangle$, where $|n_{\alpha}\rangle$ ($\alpha=x,y$) is a Fock state with $n_{\alpha}$ phonons.
\section{Phonon Adiabatic Elimination}\label{PHE}
In order to perform phonon temperature measurement by detecting the ion's internal state population we consider the spin-phonon coupling term in (\ref{Hsb}) as a perturbation which is valid as long as $|\delta_{\alpha}|\gg g$. Then the the c.m. vibrational modes in $x$-$y$ directions can be traced out which leads to an effective spin-phonon coupling \cite{Ivanov2016,Ivanov2016_1}. In the following we show that residual interaction is described by spin-dependent beam-splitter or two mode squeezing phonon operators.

Let's perform unitary transformation to (\ref{Hsb}) such that $\hat{H}_{\rm eff}=\hat{R}(\hat{H}_{\rm b}+\hat{H}_{\rm sb})\hat{R}^{\dag}$, where we set $\hat{R}=e^{-\hat{S}}$ with $\hat{S}$ being anti-Hermitian operator.
We choose $\hat{S}$ such that all terms in order of $g$ in $\hat{H}_{\rm eff}$ are canceled and the first term describing the spin-phonon interaction is of order of $g^{2}/\delta_{\alpha}$. In order to fulfill this we determine the operator $\hat{S}$ by the condition $\hat{H}_{\rm b}-[\hat{S},\hat{H}_{\rm sb}]=0$ which gives
\begin{equation}
\hat{H}_{\rm eff}\approx \hat{H}_{\rm b}+\frac{1}{2}[\hat{H}_{\rm sp},\hat{S}]+O(g^{3}/\delta_{\alpha}^{2}).
\end{equation}
Since the time evolution of $\hat{S}(t)=e^{i \hat{H}_{\rm b} t/\hbar}\hat{S}e^{-i\hat{H}_{\rm b} t/\hbar}$ is governed by the Heisenberg equation, namely $i\hbar \dot{\hat{S}}=[\hat{S}(t),\hat{H}_{\rm b}]$ we obtain $i\hbar \dot{\hat{S}}(t)=\hat{H}_{\rm sp}(t)$, where $\hat{H}_{\rm sp}(t)=e^{i \hat{H}_{\rm b} t/\hbar}\hat{H}_{\rm sp}e^{-i\hat{H}_{\rm b} t/\hbar}$. Using this we find after the integration
\begin{equation}
\hat{S}=\sum_{\alpha=x,y}\frac{2g}{\delta_{\alpha}\sqrt{N}}\hat{J}_{\alpha}(\hat{a}_{\alpha}-\hat{a}_{\alpha}^{\dag}).
\end{equation}
Then for the effective Hamiltonian we obtain
\begin{eqnarray}
\hat{H}_{\rm eff}&=&\hat{H}_{\rm b}+\sum_{\alpha=x,y}\frac{4\hbar g^{2}}{N\delta_{\alpha}}\hat{J}_{\alpha}^{2}+\frac{2i\hbar g^{2}}{N\delta_{x}\delta_{y}}\hat{J}_{z}\{(\delta_{x}+\delta_{y})(\hat{a}_{x}^{\dag}\hat{a}_{y}-\hat{a}_{x}\hat{a}_{y}^{\dag})\notag\\
&&-(\delta_{x}-\delta_{y})(\hat{a}_{x}^{\dag}\hat{a}_{y}^{\dag}-\hat{a}_{x}\hat{a}_{y})\},\label{Heff}
\end{eqnarray}
where we have omitted the constant terms. The second term in (\ref{Heff}) describes the long-range spin-spin interaction mediated by the c.m. vibrational modes. The last term in (\ref{Heff}) is the residual spin-phonon coupling, which we use to map the relevant temperature information of the c.m. vibrational modes into the spin state populations. Depending on the phonon detunings we distinguish to two cases:
\subsection{Spin-Dependent Phonon Beam-Splitter Operator}
Setting $\delta_{x}=\delta_{y}=\delta$ the expression (\ref{Heff}) simplifies to
\begin{equation}
\hat{H}_{\rm bs}=\hat{H}_{\rm b}+\frac{4 \hbar g^{2}}{N\delta}\hat{J}_{z}^{2}-\frac{4i\hbar g^{2}}{N \delta}\hat{J}_{z}(\hat{a}_{x}^{\dag}\hat{a}_{y}-\hat{a}_{x}\hat{a}_{y}^{\dag}),\label{Hbs}
\end{equation}
where we use the relation $\hat{J}_{x}^{2}+\hat{J}_{y}^{2}=\hat{J}^{2}-\hat{J}_{z}^{2}$. The Hamiltonian (\ref{Hbs}) contains a quadratic term in the collective spin operator and residual spin-phonon term, where both are diagonal in the collective spin basis. The unitary evolution generated by the residual spin-phonon term is given by the spin-dependent beam-splitter operator \cite{Campos1989}. Such a operator has been used as an entangler of the output optical fields. Here we use this spin-dependent part to map the temperature of the quantum oscillators into the collective spin states.
\subsection{Two-mode Squeezing Operator}
For $\delta_{x}=-\delta_{y}$ the Hamiltonian (\ref{Heff}) becomes
\begin{equation}
\hat{H}_{\rm tms}=\hat{H}_{\rm b}+\frac{4 \hbar g^{2}}{N\delta}(\hat{J}_{x}^{2}-\hat{J}_{y}^{2})-\frac{4i\hbar g^{2}}{N \delta}\hat{J}_{z}(\hat{a}_{x}^{\dag}\hat{a}_{y}^{\dag}-\hat{a}_{x}\hat{a}_{y}).\label{Htms}
\end{equation}
The spin part in (\ref{Htms}) is described by the Lipkin-Meshkov-Glick Hamiltonian \cite{Lipkin1965}. The unitary evolution generated by the residual spin-phonon term is the spin-dependent two-mode squeezing operator \cite{Caves1991,yi1996} in which two phonons in $x$-$y$ directions are simultaneously created/anihilated. Again we shall use this interaction to map the temperature information of the quantum oscillators into the spin-degree of freedom.

In the following we briefly provide the general background of the theory of the temperature estimation.
\section{Temperature Estimation: General Background}\label{TEGB}

\begin{figure}
\includegraphics[width=0.45\textwidth]{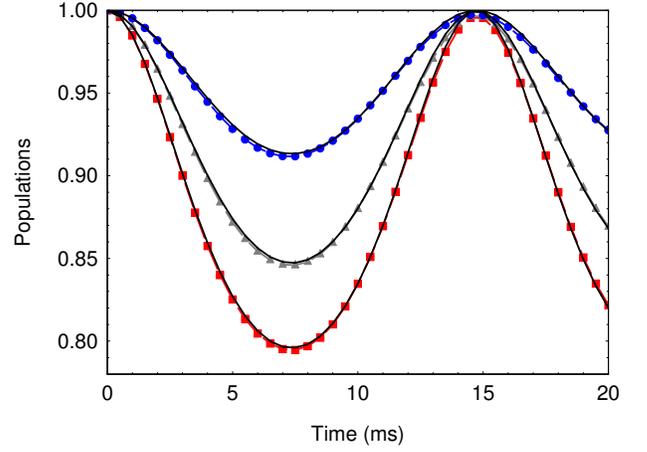}
\caption{(Color online) Time evolution of the spin state probability $p_{\uparrow}(t,T)$ for various c.m. mean-phonon numbers $\bar{n}$. We compared the numerical solution of the time-dependent Lioville equation $i\hbar\frac{d}{dt}\hat{\rho}=[\hat{H},\hat{\rho}]$ with Hamiltonian Eq. (\ref{Heff}) after applying $\pi/2$ laser pulse, with the analytical solution (black solid line) Eq. (\ref{p}). The parameters are set $g=4$ kHz, $\delta=150$ kHz, and $\phi=0$. We assume initial thermal state with mean phonon number $\bar{n}=0.15$ (dashed red square), $\bar{n}=0.1$ (dashed grey triangle), $\bar{n}=0.05$ (dashed blue dot). The blue solid line represent the analytical solution (\ref{p_y=0}) and respectively the red dashed line is the numerical solution assuming $\bar{n}_{x}=0.15$.}
\label{fig1}
\end{figure}
We consider that the two c.m. vibrational modes are in thermal state with inverse temperature $\beta=1/k_{\rm B}T$ where $k_{\rm B}$ is the Boltzmann constant and $T$ is the temperature, the parameter which we wish to estimate. The state of the two quantum oscillators is described by the Gibbs density operator
\begin{equation}
\hat{\rho}_{\rm th}=\sum_{n_{x}=0}^{\infty}\sum_{n_{y}=0}^{\infty}P_{n_{x}}P_{n_{y}}\left|n_{x},n_{y}\right\rangle\left\langle n_{x},n_{y}\right|,
\end{equation}
where $P_{s}=\frac{\bar{n}_{\alpha}^{s}}{(1+\bar{n}_{\alpha})^{s+1}}$ and $\bar{n}_{\alpha}=(e^{\beta\hbar\omega_{\alpha}}-1)^{-1}$ being the average number of thermal excitations. Usually, such a thermal state is realized experimentally after Doppler cooling of the ion crystal \cite{Wineland1979}. Following the approach discussed in \cite{Brunelli2011,Brunelli2012} the temperature estimation is performed by read out the spin state populations via state dependent fluorescence technique. For this goal we assume that the total density operator evolves in time according to $\hat{\rho}(t)=\hat{U}\hat{\rho}(0)\hat{U}^{\dag}$, where $\hat{\rho}(0)=\hat{\rho}_{\rm spin}(0)\otimes\hat{\rho}_{\rm th}$ is the initial density operator with $\hat{\rho}_{\rm spin}(0)$ being the initial spin density operator and $\hat{U}=e^{-i \hat{H}_{\rm eff} t/\hbar}$ is the unitary operator. At time $t$ the spin density operator is $\hat{\rho}_{\rm spin}(t)={\rm Tr}_{\rm p}(\hat{\rho})$ where the tracing over the phonon degree of freedom is performed. For a set of measurement outcomes with probability $p_{m}(T)$ with $m=-j,\ldots,j$, the classical Fisher information quantifies the amount of information on the temperature of the system. We have
\begin{equation}
F_{\rm CL}(T)=\sum_{m=-j}^{j}\frac{\left(\partial_{T}p_{m}\right)^{2}}{p_{m}}.\label{Fcl}
\end{equation}
The Cramer-Rao inequality bounded the variance of the temperature estimation
\begin{equation}
\delta T^{2}\geq\frac{1}{\nu F_{\rm CL}(T)},\label{ccrb}
\end{equation}
where $\nu$ is the number of experimental repetitions. The classical Fisher information is further bounded by the quantum Fisher information $F_{\rm Q}(T)$ which gives the ultimate limit of precision in the temperature estimation quantified by the quantum Cramer-Rao bound
\begin{equation}
\delta T^{2}\geq\frac{1}{\nu F_{\rm Q}(T)}\label{qcrb}.
\end{equation}
The quantum Fisher information can be expressed as $F_{\rm Q}(T)={\rm Tr}(\hat{\rho}_{\rm spin}\hat{L}^{2})$, where $\hat{L}(T)$ is the symmetrical logarithmic derivative operator which satisfy the operator equation $\partial_{T}\hat{\rho}_{\rm spin}=(\hat{\rho}_{\rm spin}\hat{L}+\hat{L}\hat{\rho}_{\rm spin})/2$. In present context the quantum Fisher information is a measure of distinguishability of two quantum states with respect to the infinitesimal variation of the temperature \cite{Braunstein1994}.

Finally, one can express the quantum Fisher information in the eigenbasis of the density operator $\hat{\rho}_{\rm spin}=\sum_{m=-j}^{j}\rho_{m}\left|\psi_{m}\right\rangle\left\langle\psi_{m}\right|$ where $\rho_{m}$ and $\left|\psi_{m}\right\rangle$ are respectively the $m$th eigenvalue and eigenvector. We have \cite{Paris2009,Pezze2014} (see Appendix \ref{QFI})
\begin{eqnarray}
&&F_{\rm Q}(T)=F_{\rm Q}^{\rm cl}(T)+F_{\rm Q}^{\rm nc}(T),\quad F_{\rm Q}^{\rm cl}(T)\sum_{m=-j}^{j}\frac{(\partial_{T}\rho_{m})^{2}}{\rho_{m}},\notag\\
&&F_{\rm Q}^{\rm nc}(T)=2\sum_{m\neq k}^{j}\frac{(\rho_{m}-\rho_{k})^{2}}{\rho_{m}+\rho_{k}}
|\langle\partial_{T}\psi_{m}|\psi_{k}\rangle|^{2},\label{QFI_Express}
\end{eqnarray}
The first term in (\ref{QFI_Express}) represent the classical Fisher information for the probability distribution $\rho_{m}$ while the second term has truly quantum contribution and leads to $F_{\rm Q}(T)\geq F_{\rm CL}(T)$.
\begin{figure}
\includegraphics[width=0.45\textwidth]{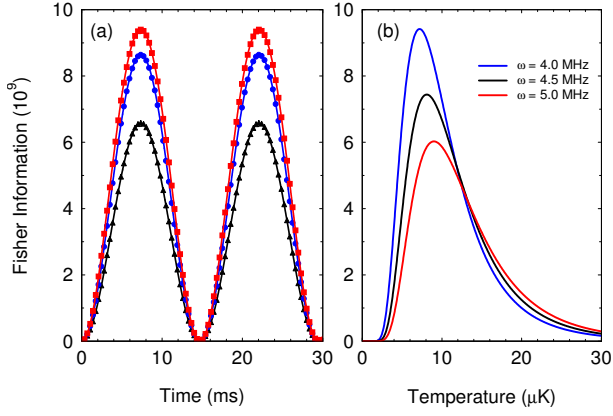}
\caption{(Color online) (a) Classical Fisher information as a function of time. We assume ion trap frequency $\omega=4$ MHz and temperature $T=5$ $\mu$K (black triangles), $T=6$ $\mu$K (blue dots), $T=7$ $\mu$K (red squares). (b) $F_{\rm CL}$ at $t_{\rm max}$ for different trap frequencies $\omega$.}
\label{fig2}
\end{figure}

\section{Temperature estimation with beam-splitter phonon operator}\label{TEBSO}
\subsection{Single Ion Case, $j=\frac{1}{2}$}
We begin with the case $j=\frac{1}{2}$ and assume equal trap frequencies $\omega_{x}=\omega_{y}=\omega$ which implies that the c.m. mean phonon numbers along the two orthogonal directions are equal, $\bar{n}_{x}=\bar{n}_{y}=\bar{n}$. For that case it is convenient to introduce a pair of right and left chiral operators according to $\hat{a}_{\rm r}=(\hat{a}_{x}-i\hat{a}_{y})/\sqrt{2}$ and $\hat{a}_{\rm l}=(\hat{a}_{x}+i\hat{a}_{y})/\sqrt{2}$ which can be used to diagonalized the phonon part in Eq. (\ref{Hbs}). Indeed omitting the constant term we find $\hat{H}_{\rm bs}=\frac{2\hbar g^{2}}{\delta}(\hat{n}_{\rm l}-\hat{n}_{\rm r})\sigma_{z}$.
\begin{figure}
\includegraphics[width=0.45\textwidth]{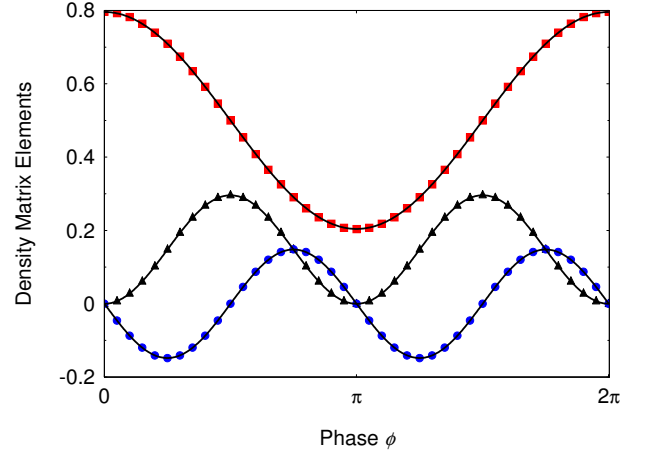}
\caption{(Color online) Matrix elements of the spin density operator $\hat{\rho}_{\rm spin}$ at time $t_{\rm max}$ versus the laser phase $\phi$. We assume mean-phonon number $\bar{n}=0.15$. We compared the exact results for the spin population $p_{\uparrow}(t_{\rm max},T)$ (red squares) and the spin coherence $\Re p_{\uparrow,\downarrow}(t_{\rm max},T)$ (black triangles), $\Im p_{\uparrow,\downarrow}(t_{\rm max},T)$ (blue dots) with the analytical expressions (black solid lines), Eqs. (\ref{p}) and (\ref{coherence}).}
\label{fig3}
\end{figure}

The Ramsey-type sequence starts by preparing the spin in the equal superposition, $\hat{\rho}_{\rm spin}(0)=(\left|\uparrow\right\rangle+\left|\downarrow\right\rangle)(\left\langle\uparrow\right|+\left\langle\downarrow\right|)/2$. The combined system evolves for time $t^{\prime}$ according the beam-splitter unitary operator $\hat{U}_{\rm bs}=e^{-i \hat{H}_{\rm bs}t^{\prime}/\hbar}$. Then a $\frac{\pi}{2}$ laser pulse with phase $\phi$ is applied which creates the spin superposition, $\left|\uparrow\right\rangle\rightarrow (\left|\uparrow\right\rangle-e^{-i\phi}\left|\downarrow\right\rangle)/\sqrt{2}$ and $\left|\downarrow\right\rangle\rightarrow (\left|\downarrow\right\rangle+e^{i\phi}\left|\uparrow\right\rangle)/\sqrt{2}$ which conclude the interaction sequence. The spin density operator at time $t$ becomes $\hat{\rho}_{\rm spin}(t,T)=p_{\uparrow}\left|\uparrow\right\rangle\left\langle\uparrow\right|+p_{\downarrow}\left|\downarrow\right\rangle\left\langle\downarrow\right|
+p_{\uparrow,\downarrow}\left|\uparrow\right\rangle\left\langle\downarrow\right|
+p_{\downarrow,\uparrow}\left|\downarrow\right\rangle\left\langle\uparrow\right|$. Subsequently a measurement of the spin population is performed with probability given by
\begin{equation}
p_{\uparrow}(t,T,\phi)=\frac{1}{2}\left(1+\frac{\cos(\phi)}{1+4\bar{n}(\bar{n}+1)\sin^{2}\left(\frac{\theta t}{2}\right)}\right),\label{p}
\end{equation}
with $p_{\downarrow}=1-p_{\uparrow}$ and $\theta=\frac{4g^{2}}{\delta}$. The off-diagonal quantum coherence elements are
\begin{equation}
p_{\uparrow,\downarrow}(t,T,\phi)=-\frac{i}{2}\frac{e^{i\phi}\sin(\phi)}{1+4\bar{n}(\bar{n}+1)\sin^{2}\left(\frac{\theta t}{2}\right)},\label{coherence}
\end{equation}
with $p_{\downarrow,\uparrow}=(p_{\uparrow,\downarrow})^{*}$
\begin{figure}
\includegraphics[width=0.45\textwidth]{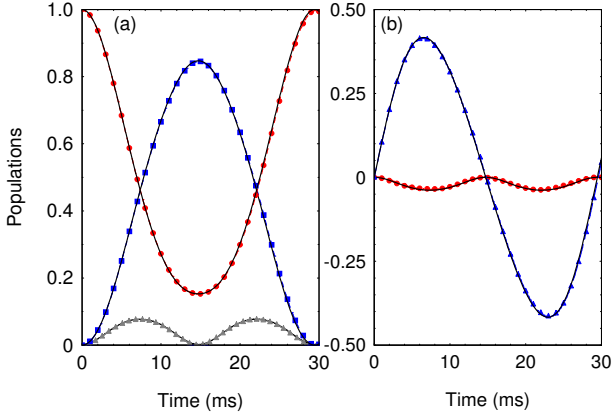}
\caption{(Color online).(a) Populations $p_{m}(t,T)$ ($m=-1,0,1$) for an ion chain with two ions as a function of time. We compared numerical solution with Hamiltonian (\ref{Hsb}) for $p_{1}(t,T)$ (dashed red dots), $p_{0}(t,T)$ (dashed grey triangles), $p_{-1}(t,T)$ (dashed blue squares) with the analytical expression, Eq. (\ref{pop_j=1}) (black solid lines). We assume mean phonon number $\bar{n}=0.1$. (b) Off-diagonal coherence element $p_{1,-1}(t,T)$ of the spin-density matrix. We compared the exact results for $\Im p_{1,-1}$ (dashed blue triangles), $\Re p_{1,-1}$ (dashed red dots) with Eq. (\ref{coh_j=1}) (black solid lines).}
\label{fig4}
\end{figure}

In Fig. (\ref{fig1}) we show the time evolution of the probability $p_{\uparrow}(t,T)$ for different $\bar{n}$. We observe very good agreement between the analytical expression (\ref{p}) and the exact result with Hamiltonian (\ref{Heff}). As can be seen from the figure the signal oscillation period vary with $\bar{n}$ and approaches to zero in the limit $T\rightarrow 0$. Using Eqs. (\ref{Fcl}) and (\ref{p}) one can obtain the classical Fisher information. We find for $\phi=0$
\begin{equation}
F_{\rm CL}=\frac{\hbar^{2}\omega^{2}\sin^{2}\left(\frac{\theta t}{2}\right)\sinh^{2}\left(\frac{\hbar\omega}{k_{\rm B}T}\right)}{k_{\rm B}^{2}T^{4}(\cosh\left(\frac{\hbar\omega}{k_{\rm B}T}\right)-\cos
\left(\frac{\theta t}{2}\right))(\cosh\left(\frac{\hbar\omega}{k_{\rm B}T}\right)-\cos(\theta t))^{2}}.\label{clf}
\end{equation}
In Fig. \ref{fig2}(a) we plot the classical Fisher information (\ref{clf}) versus the interaction time $t$. At low temperature $T k_{\rm B}\leq \hbar\omega/2$ the maximum value of $F_{\rm CL}(T)$ is reached at time $\theta t_{\rm max}=(2k+1)\pi$, ($k=0,1,2,\ldots$) and $\phi=2p\pi$ ($p=0,1,2,\ldots$). In Fig. \ref{fig2}(b) we show $F_{\rm CL}(T)$ at $t_{\rm max}$ versus the temperature $T$. It has a maximal value defined by the condition $\partial_{T}F_{\rm CL}(T)=0$ which reduces to the following transcendental equation $x-x{\rm sech}(x)(2+{\rm sech}(x))-4\tanh(x)=0$ where $x=\beta\hbar\omega$.  We find that the maximal value of the classical Fisher information is achieved for temperature $T_{\rm max}\approx \frac{\hbar\omega}{4.245k_{\rm B}}$. At this point the uncertainty in the estimation of the temperature is bounded by the classical Cramer-Rao bound (\ref{ccrb}) which yields $\delta T\geq\frac{\hbar\omega}{2.964\sqrt{\nu}k_{\rm B}}$.
As an example consider transverse ion trap frequency $\omega=2$ MHz we obtain temperature sensitivity approximately to $5.2$ $\mu$K.

In Fig. \ref{fig3} we plot the spin density matrix elements at the time $t_{\rm max}$ as a function of the laser phase $\phi$. We see from Eqs. (\ref{p}) and (\ref{coherence}) that the spin density operator becomes diagonal for laser phase $\phi=2p\pi$ with $p$ integer, leading to the equality $F_{\rm CL}(T)=F_{\rm Q}(T)$ at any instance of time $t$. Indeed as long as the eigenvectors of $\hat{\rho}_{\rm spin}(t,T)$ do not depend on the temperature such a equality is always fulfilled, see Eq. (\ref{QFI_Express}) and the Appendix \ref{QFI} for details. The latter implies that uncertainty of the temperature estimation performed by the projective measurements in the original spin basis is bounded by the quantum Cramer-Rao inequality (\ref{qcrb}). This result can be generalized for an arbitrary phase $\phi$ where one can find a basis $\left|\psi_{\uparrow}\right\rangle=-i e^{i\phi}\cos(\phi/2)\left|\uparrow\right\rangle+\sin(\phi/2)\left|\downarrow\right\rangle$ and $\left|\psi_{\downarrow}\right\rangle=i e^{i\phi}\sin(\phi/2)\left|\uparrow\right\rangle+\cos(\phi/2)\left|\downarrow\right\rangle$ independent on the temperature $T$ which diagonalize $\hat{\rho}_{\rm spin}(t,T)$ with eigenvalues $\rho_{\uparrow}(t,T)=p_{\uparrow}(t,T,\phi=0)$ and $\rho_{\downarrow}(t,T)=p_{\downarrow}(t,T,\phi=0)$.
\begin{figure}
\includegraphics[width=0.45\textwidth]{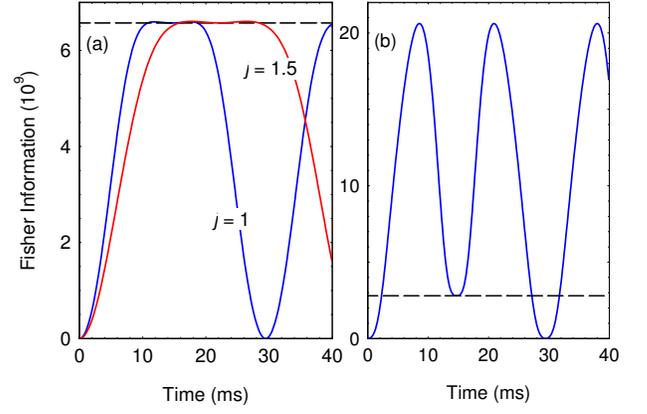}
\caption{(Color online) (a) Fisher information versus interaction time $t$. We plot the QFI for $j=1$ and $F_{\rm Q}^{\rm nc}(T)$ for $j=3/2$ at $T=5$ $\mu$K. As a comparison is shown the QFI for $j=1/2$ at $t_{\rm max}$ (black dashed lime) according Eq. (\ref{clf}). (b) QFI for $j=1$ compared to $j=1/2$ at $t_{\rm max}$ (black dashed lime) for $T=15$ $\mu$K.}
\label{fig5}
\end{figure}

Up to know we assume that both transverse vibrational modes are in thermal state with mean phonon number $\bar{n}$. Consider that one of the modes is prepared in the motion ground state for example the c.m. mode along the $y$-direction by using red-sideband laser cooling technique. To goal is to estimate the temperature of the c.m. mode along the $x$-direction which is prepared in the thermal state with mean phonon number $\bar{n}_{x}$. The vibrational density operator becomes $\hat{\rho}_{\rm th}=\sum_{n_{x}=0}^{\infty}P_{n_{x}}|n_{x},0_{y}\rangle\langle n_{x},0_{y}|$. Assuming that the spins are prepared in equal superposition with $\hat{\rho}_{\rm spin}(0)=(\left|\uparrow\right\rangle+\left|\downarrow\right\rangle)(\left\langle\uparrow\right|+\left\langle\downarrow\right|)/2$, the evolution of the total system is driven by the propagator $\hat{U}_{\rm bs}=e^{-i \hat{H}_{\rm bs}t^{\prime}/\hbar}$. After applying $\pi/2$ laser pulse (we set for simplicity $\phi=0$) the spin density operator becomes $\hat{\rho}_{\rm spin}(t,T)=p_{\uparrow}(t,T)\left|\uparrow\right\rangle\left\langle\uparrow\right|
+p_{\downarrow}(t,T)\left|\downarrow\right\rangle\left\langle\downarrow\right|$ where (see Appendix \ref{SP} for details)
\begin{equation}
p_{\uparrow}(t,T)=\frac{1}{2}\left(1+\frac{1}{1+2\bar{n}_{x}\sin^{2}\left(\frac{\theta t}{2}\right)}\right)\label{p_y=0}.
\end{equation}
In Fig. (\ref{fig1}) we compare the analytical expression (\ref{p_y=0}) with the exact result. As can be seen the population oscillates with the same period as (\ref{p}) but the amplitude is less sensitive to the change of the mean phonon number. Using (\ref{p_y=0}) we obtain the classical Fisher information
\begin{equation}
F_{\rm CL}=\frac{\hbar^{2}\omega_{x}^{2}\sin^{2}\left(\frac{\theta t}{2}\right)e^{\frac{2\hbar\omega_{x}}{k_{\rm B}T}}}{k_{\rm B}^{2}T^{4}\left(e^{\frac{\hbar\omega_{x}}{k_{\rm B}T}}+\cos^{2}\left(\frac{\theta t}{2}\right)\right)\left(e^{\frac{\hbar\omega_{x}}{k_{\rm B}T}}-\cos(\theta t)\right)^{2}},
\end{equation}
with the equality $F_{\rm CL}(T)=F_{\rm Q}(T)$. The maximum of $F_{\rm CL}(T)$ is achieved for $T\approx \frac{\hbar\omega_{x}}{4.13 k_{\rm B}}$ where we obtain temperature sensitivity to $\delta T\geq (\hbar\omega/2.13\sqrt{\nu}k_{\rm B})$.
\subsection{Multi-Ion Case $j>\frac{1}{2}$}
In the following we discuss the temperature estimation of the c.m. mode using ion crystal consisting of $N$ ions. In that case the bichromatic laser-ion interaction couples the collective spin states to the c.m. vibrational mode according Eq. (\ref{Hbs}). We assume that initially the spins are fully polarized along the $x$-direction, $\left|\psi_{\rm spin}(0)\right\rangle=\left|\uparrow\uparrow\ldots\uparrow\right\rangle_{x}$. Using the collective Dicke states one can express the initial spin state as
\begin{equation}
\left|\psi_{\rm spin}(0)\right\rangle=\sum_{m=-j}^{j}\sqrt{\frac{(2j)!}{2^{2j}(j+m)!(j-m)!}}\left|j,m\right\rangle.
\end{equation}
The combine system evolves in time according the unitary operator $\hat{U}_{\rm bs}=e^{-i \hat{H}_{\rm bs} t^{\prime}/\hbar}$. Similar to a single ion case at time $t^{\prime}$ a global $\pi/2$ laser pulse is applied to all spins which conclude the interaction sequence.
\begin{figure}
\includegraphics[width=0.45\textwidth]{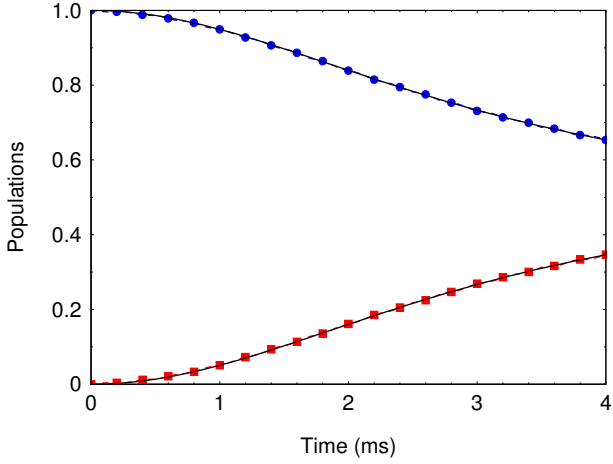}
\caption{(Color online) Time evolution of the spin populations. The detuning is set to $|\delta|=150$ kHz and the system evolves according the spin-dependent two-mode squeezing operator. We compare the exact numerical solution for $p_{\uparrow}(t,T)$ (dashed blue dots) and $p_{\downarrow}(t,T)$ (dashed red squares) with the analytical formulas (black solid lines), Eq. (\ref{pop_sq}). We assume mean-phonon number $\bar{n}=0.1$. }
\label{fig6}
\end{figure}

Consider as an example ion chain with two ions, $j=1$. The spin density operator at time $t$ becomes $\hat{\rho}_{\rm spin}(t,T)=\sum_{m=-1}^{1}p_{m}|j,m\rangle\langle m,j|+\{p_{1,-1}|1,1\rangle\langle-1,1|+{\rm h.c.}\}$, with populations
\begin{eqnarray}
p_{\pm 1}(t,T)&=&\frac{1}{8}\{3+\frac{1}{1+4\bar{n}(\bar{n}+1)\sin^{2}\left(\frac{\theta t}{2}\right)}\notag\\
&&\pm\frac{4\cos\left(\frac{\theta t}{2}\right)}{1+4\bar{n}(\bar{n}+1)\sin^{2}\left(\frac{\theta t}{4}\right)}\},\label{pop_j=1}
\end{eqnarray}
$p_{0}=1-p_{1}-p_{-1}$ and quantum coherence
\begin{eqnarray}
p_{1,-1}(t,T)&=&\frac{1}{8}\{-1+\frac{1}{1+4\bar{n}(\bar{n}+1)\sin^{2}\left(\frac{\theta t}{2}\right)}\notag\\
&&+\frac{4i\sin\left(\frac{\theta t}{2}\right)}{1+4\bar{n}(\bar{n}+1)\sin^{2}\left(\frac{\theta t}{4}\right)}\}.\label{coh_j=1}
\end{eqnarray}
Figure (\ref{fig4}) compares the analytical formulas (\ref{pop_j=1}) and (\ref{coh_j=1}) to the exact solution where very good agreement is observed. In order to find the quantum Fisher information we diagonalize the spin density operator $\hat{\rho}_{\rm spin}(t,T)$, see Appendix \ref{SP}. In contrast to a single ion case, now the non-classical part $F_{\rm Q}^{\rm nc}(T)$ of the quantum Fisher information is generally non-zero leading to $F_{\rm Q}(T)\ge F_{\rm CL}(T)$ for $j>1/2$. In Fig. \ref{fig5} we plot the Fisher information versus the interaction time $t$. We observe that at low temperature limit $T\rightarrow 0$ the eigenvectors of $\rho_{\rm spin}(t,T)$ becomes temperature independent such that non-classical part of QFI for $j=1$ tends to zero, which leads to equality between the classical and quantum Fisher information. In this regime the QFI becomes approximately equal to the maximal value of QFI for $j=1/2$. However, in contrast to the single ion case now we find plateaus where the maximal value of QFI is reached. This could be experimental advantage since it does not require precise knowledge of the parameter $\theta$ as is the case for $j=1/2$ where the maximal sensitivity is achieved for $t_{\rm max}=\pi/\theta$. As can be seen from Fig. \ref{fig5}a the size of the plateaus increases with the number of ions. On the other hand slightly rising the temperature leads to a non vanishing $F_{\rm Q}^{\rm nc}(T)$ such that the QFI for $j>1/2$ becomes much higher than the maximal value of QFI for $j=1/2$, as is shown in Fig. \ref{fig5}b.
\section{Temperature estimation with two mode squeezing operator}\label{TETMSO}
\begin{figure}
\includegraphics[width=0.45\textwidth]{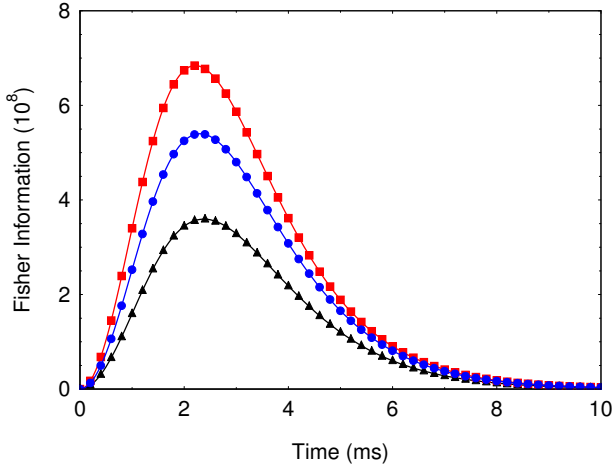}
\caption{(Color online) Classical Fisher Information versus the interaction time for different temperatures. We calculate $F_{\rm CL}(T)$ using formula (\ref{pop_sq}) for $T=5$ $\mu$K (black triangles), $T=6$ $\mu$K (blue triangles) and $T=7$ $\mu$K (red squares) assuming trap frequency $\omega=4$ MHz.    }
\label{fig7}
\end{figure}
In the following we discuss the temperature estimation using two-mode squeezing Hamiltonian, Eq. (\ref{Htms}). We consider only the single ion case such that the first term in Eq. (\ref{Htms}) has no role since it gives constant shift. Again, following the Ramsey sequence the initial prepared equal spin superposition evolves in time according to two-mode squeezing transformation $\hat{U}_{\rm tms}(t)=e^{-i \hat{H}_{\rm tms}t/\hbar}$. At time $t^{\prime}$ a $\pi/2$ laser pulse is applied and subsequently the temperature estimation is carried out by state projective measurements.
The analytical expression for the spin population can be derived exactly, see Appendix \ref{SP}. We compare these formulas with the exact numerical result. From the Fig. \ref{fig6} we see that the expressions (\ref{pop_sq}) matches the exact result very closely. We find that for sufficiently long interaction time both signals tend to $1/2$ such that the spin density operator becomes completely incoherent mixture and no temperature estimation is possible. Also at any instance of time the off-diagonal elements of $\rho_{\rm spin}(t,T)$ are zero such that the density operator is diagonal in the original spin basis. The latter implies that we need only to consider $F_{\rm CL}(T)$ because of the equality $F_{\rm CL}(T)=F_{\rm Q}(T)$. Using the expression (\ref{pop_sq}) one can calculate the classical Fisher information. On one hand we observe that the maximal value of $F_{\rm CL}(T)$ using two-mode squeezing transformation is smaller than the corresponding value using beam-splitter operator, see Eq. (\ref{clf}) and Fig. \ref{fig2}a for comparison. On the other hand the maximal value of $F_{\rm CL}(T)$ is reached for shorter interaction time which could have particular advantage in case of strong motion heating which reduces the coherence time.

\section{Conclusion}\label{DO}
We have considered the quantum estimation scheme of the temperature of the transverse vibrational modes of linear ion chain. The protocol is a Ramsey type sequence in which the acquired phase of the spins is mapped on the respective population which subsequently is measured. We characterize the temperature sensitivity in terms of classical and quantum Fisher information and show that scheme using single ion as a quantum probe is optimal in a sense that the state projective measurement saturates the Cramer-Rao bound. We find a measurement basis which leads to equality between the classical and quantum Fisher information and show that is independent on the temperature. At low temperature we find plateau of maximal sensitivity in the case of multi-ion quantum probe. We have shown that the size of the plateau increases with the number of ions. Rising the temperature we have shown that non-classical part of the quantum Fisher information leads to enhancement of the temperature sensitivity compared with the single ion case.

\section*{Acknowledgments}
PAI acknowledges support by the ERyQSenS, Bulgarian Science Fund Grant No. DO02/3.
\appendix
\section{Quantum Fisher Information}\label{QFI}
Consider the quantum Fisher information $F_{\rm Q}(T)=\sum_{m=-j}^{j}\rho_{m}\langle\psi_{m}|\hat{L}^{2}|\psi_{m}\rangle$ which gives the ultimate precision in the temperature estimation. Here $\hat{L}$ is the symmetric logarithmic derivative operator which satisfies the operator equation $\partial_{T}\hat{\rho}_{\rm spin}=(\hat{\rho}_{\rm spin}\hat{L}+\hat{L}\hat{\rho}_{\rm spin})/2$. We decompose the spin density operator in its eigenbasis, namely $\hat{\rho}_{\rm spin}=\sum_{m=-j}^{j}\rho_{m}|\psi_{m}\rangle\langle\psi_{m}|$ where $\rho_{m}$ and $|\psi_{m}\rangle$ are respectively the eigenvalue and the corresponding eigenvector. The symmetric logarithmic derivative operator can be written as
\begin{eqnarray}
&&\hat{L}=2\sum_{m,k=-j}^{j}\frac{\langle\psi_{m}|\partial_{T}\hat{\rho}_{\rm spin}|\psi_{k}\rangle}{p_{m}+p_{k}}|\psi_{m}\rangle\langle\psi_{k}|\notag\\
&&=\sum_{m=-j}^{j}\frac{\partial_{T} \rho_{m}}{\rho_{m}}|\psi_{m}\rangle\langle\psi_{m}|
+2\sum_{m\neq k}^{j}\frac{\langle\psi_{m}|\partial_{T}\hat{\rho}_{\rm spin}|\psi_{k}\rangle}{\rho_{m}+\rho_{k}}|\psi_{m}\rangle\langle\psi_{k}|\notag\\
&&=\sum_{m=-j}^{j}\frac{\partial_{T} \rho_{m}}{\rho_{m}}|\psi_{m}\rangle\langle\psi_{m}|+2\sum_{m\neq k}^{j}\frac{\rho_{m}-\rho_{k}}{\rho_{m}+\rho_{k}}\langle\partial_{T}\psi_{m}|\psi_{k}\rangle|\psi_{m}\rangle\langle\psi_{k}|.\label{L}
\end{eqnarray}
Here we have used the orthogonality $\langle\psi_{m}|\psi_{k}\rangle=\delta_{m,k}$ as well as the relation $\langle\partial_{T}\psi_{m}|\psi_{k}\rangle=-\langle\psi_{m}|\partial_{T}\psi_{k}\rangle$. Finally, using Eq. (\ref{L}) one can write the quantum Fisher information in the following way
\begin{equation}
F_{\rm Q}(t,T)=\sum_{m=-j}^{j}\frac{(\partial_{T}\rho_{m})^{2}}{\rho_{m}}+2\sum_{m\neq k}^{j}\frac{(\rho_{m}-\rho_{k})^{2}}{\rho_{m}+\rho_{k}}
|\langle\partial_{T}\psi_{m}|\psi_{k}\rangle|^{2},\label{qfi_N}
\end{equation}
where the first term is the classical part and the second term has respectively truly quantum contribution.
\section{Spin Populations}\label{SP}
\subsection{Beam-Splitter Operator}
The time evolution of the spin-density operator after taking the partial trace over the vibrational degrees of freedom is
\begin{equation}
\hat{\rho}_{\rm spin}=\sum_{n,s=0}^{\infty}P_{n}P_{s}e^{-i\frac{\theta t}{N}(\hat{J}_{z}^{2}+\hat{J}_{z}(n-s))}
\hat{\rho}_{\rm s}(0)e^{i\frac{\theta t}{N}(\hat{J}_{z}^{2}+\hat{J}_{z}(n-s))},
\end{equation}
where $P_{s}=\frac{\bar{n}^{s}}{(1+\bar{n})^{s+1}}$. We assume that the initial spin density operator is $\hat{\rho}_{\rm s}(0)=|\psi(0)\rangle\langle\psi(0)|$ with all spins polarized along the $x$-axis, $\left|\psi(0)\right\rangle=\left|\uparrow\uparrow\dots\uparrow\right\rangle_{x}$. At time $t^{\prime}$ a global $\pi/2$ laser pulse is applied to all ions which rotates the spin states $\left|\uparrow\right\rangle\rightarrow (\left|\uparrow\right\rangle-e^{-i\phi}\left|\downarrow\right\rangle)/\sqrt{2}$ and $\left|\downarrow\right\rangle\rightarrow (\left|\downarrow\right\rangle+e^{i\phi}\left|\uparrow\right\rangle)/\sqrt{2}$ with laser phase $\phi$. Measuring the spin populations allow to extract the information of the temperature of the c.m. vibrational mode.

For particular example of single ion system with $j=1/2$ we obtain the following density operator $\hat{\rho}_{\rm spin}(t,T)=p_{\uparrow}\left|\uparrow\right\rangle\left\langle\uparrow\right|+p_{\downarrow}\left|\downarrow\right\rangle\left\langle\downarrow\right|
+p_{\uparrow,\downarrow}\left|\uparrow\right\rangle\left\langle\downarrow\right|+
p_{\downarrow,\uparrow}\left|\downarrow\right\rangle\left\langle\uparrow\right|$, where
\begin{eqnarray}
&&p_{\uparrow}(t,T)=\frac{1}{2}\left(1+\sum_{n,s=0}^{\infty}P_{n}P_{s}\cos\left(\theta t(n-s)\right)\right),\notag\\
&&p_{\uparrow,\downarrow}(t,T)=i\sum_{n,s=0}^{\infty}P_{n}P_{s}\sin\left(\theta t(n-s)\right),\label{popul}
\end{eqnarray}
with $p_{\downarrow}=1-p_{\uparrow}$, $p_{\downarrow,\uparrow}=(p_{\uparrow,\downarrow})^{*}$ and $\phi=0$. Both sums in Eq. (\ref{popul}) can be evaluated which gives respectively Eq. (\ref{p}) and $p_{\uparrow,\downarrow}=0$. Thus, the density operator is diagonal with eigenstates $\left|\psi_{-1/2}\right\rangle=\left|\downarrow\right\rangle$, $\left|\psi_{1/2}\right\rangle=\left|\uparrow\right\rangle$. From Eq. (\ref{qfi_N}) we find the equality $F_{\rm CL}(t,T)=F_{\rm Q}(t,T)$. Note that this result can be generalized for arbitrary phase $\phi$.

In order to calculate the probability (\ref{p_y=0}) we use the following relation
\begin{eqnarray}
&&{\rm Tr}\{e^{\theta t(\hat{a}^{\dag}_{x}\hat{a}_{y}-\hat{a}_{x}\hat{a}^{\dag}_{y})}\hat{\rho}_{\rm th,x}\otimes|0_{y}\rangle\langle 0_{y}|\}\notag\\
&&=\sum_{n_{x}=0}^{\infty}P_{n_{x}}\langle n_{x},0_{y}|e^{\theta t(\hat{a}^{\dag}_{x}\hat{a}_{y}-\hat{a}_{x}\hat{a}^{\dag}_{y})}|n_{x},0_{y}\rangle
=\sum_{n_{x}=0}^{\infty}P_{n_{x}}\cos^{n_{x}}(\theta t)\notag\\
&&=\frac{1}{1+2\bar{n}_{x}\sin^{2}\left(\frac{\theta t}{2}\right)}.
\end{eqnarray}

We continue with the two ion case, where the collective spin populations and the spin coherences are given by Eqs. (\ref{pop_j=1}) and (\ref{coh_j=1}). We diagonalize the spin density operator which yield $\hat{\rho}_{\rm spin}(t,T)=\sum_{m=-1}^{1}\rho_{m}\left|\psi_{m}\right\rangle\left\langle\psi_{m}\right|$ where the eigenvectors are
\begin{eqnarray}
&&|\psi_{1}\rangle=e^{i\varphi}\cos(\xi)|1,1\rangle+\sin(\xi)|1,-1\rangle,\quad |\psi_{0}\rangle=|1,0\rangle,\notag\\
&&|\psi_{-1}\rangle=-e^{i\varphi}\sin(\xi)|1,1\rangle+\cos(\xi)|1,-1\rangle,
\end{eqnarray}
with
\begin{eqnarray}
&&\xi=\arctan\left(\frac{\sqrt{(a-1)^{2}+16b^{2}\sin^{2}\left(\frac{\theta t}{2}\right)}}{\sqrt{(a-1)^{2}+16b^{2}}-4b\cos\left(\frac{\theta
t}{2}\right)}\right),\notag\\
&&\varphi=\arctan\left(\frac{4b\sin\left(\frac{\theta t}{2}\right)}{a-1}\right).
\end{eqnarray}
The corresponding eigenvalues are given by
\begin{eqnarray}
&&\rho_{\pm}=\frac{1}{8}\left(3+a\pm\sqrt{(a-1)^{2}+16 b^{2}}\right),\notag\\
&&\rho_{0}=\frac{1}{4}(1-a).
\end{eqnarray}
Here we have introduced the notation $a=(1+4\bar{n}(\bar{n}+1)\sin^{2}(\frac{\theta t}{2}))^{-1}$ and $b=(1+4\bar{n}(\bar{n}+1)\sin^{2}(\frac{\theta t}{4}))^{-1}$.
\subsection{Two-mode squeezing transformation}
Using the representation of the two-mode squeezing operator presented in \cite{yi1996} one can derived expression for the spin populations. We find
\begin{eqnarray}
p_{\uparrow}(t,T)&=&\frac{1}{2}\{1+\sum_{n,s=0}^{\infty}\sum_{l,k=0}^{\min(n,s)}P_{n}P_{s}{\rm sech}(\theta t)^{n+s-l-k+1}e^{\theta t(l-k)}\notag\\
&&\times\frac{(n+s-l-k)!n!s!}{l!(n-l)!(s-l)!k!(n-k)!(s-k)!}\}\label{pop_sq}
\end{eqnarray}
and $p_{\downarrow}(t,T)=1-p_{\uparrow}(t,T)$. The off-diagonal elements of $\rho_{\rm spin}(t,T)$ are zero.


\begin{thebibliography}{99}

\bibitem{Klinkert2009} B. Kinkert and F. Narberhaus, Cell. Mol. Life Sci. \textbf{66}, 2661 (2009).

\bibitem{Gemmer2004} J. Gemmer, M. Michel, and G. Mahler, \emph{Quantum Thermodynamics} (Springer, Berlin, 2004).

\bibitem{An2014} S. An, J.-N. Zhang, M. Um, D. Lv, Y. Lu, J. Zhang, Z.-Q. Yin, H. T. Quan, and K. Kim, Nat. Phys. \textbf{11}, 193 (2015).

\bibitem{Rosnagel2016} J. Ro{\ss}nagel, S. T. Dawkins, K. N. Tolazzi, O. Abah, E. Lutz, F. Schmidt-Kaler, and K. Singer, Science \textbf{352}, 325 (2016).

\bibitem{Stace2010} T. M. Stace, Phys. Rev. A \textbf{82}, 011611(R) (2010).

\bibitem{Paris2016} M. G. A. Paris, J. Phys. A: Math. Theor. \textbf{49}, 03LT02 (2016).

\bibitem{Correa2015} L. A. Correa, M. Mehboudi, G. Adesso, and A. Sanpera, Phys. Rev. Lett. \textbf{114}, 220405 (2015).

\bibitem{Brunelli2011} M. Brunelli, S. Ovivares, and M. G. A. Paris, Phys. Rev. A \textbf{84}, 032105 (2011).

\bibitem{Brunelli2012} M. Brunelli, S. Ovivares, M. Paternistro, and M. G. A. Paris, Phys. Rev. A \textbf{86}, 012125 (2012).

\bibitem{Higgins2012} K. D. B. Higgins, B. W. Lovett, and E. M. Gauger, Phys. Rev. B \textbf{88}, 155409 (2013).

\bibitem{Sabin2014} C. Sabin, A. White, L. Hackermuller, and I. Fuentes, Sci. Rep. \textbf{4}, 6436 (2014).

\bibitem{Mehboundi2015} M. Mehboudi, M. Moreno-Cardoner, G. De Chiara, and A. Sanpera. New J. Phys. \textbf{17}, 055020 (2015).

\bibitem{Turchette2000} Q. A. Turchette, D. Kielpinski, B. E. King, D. Leibfried, D. M. Meekhof, C. J. Myatt, M. A. Rowe, C. A. Sackett, C. S. Wood, W. M. Itano, C. Monroe, and D. J. Wineland, Phys. Rev. A \textbf{61}, 063418 (2000).

\bibitem{Singer2015} J. Ro{\ss}nagel, K. N. Tolazzi, F. Schmidt-Kaler, and K. Singer, New J. Phys. \textbf{17}, 045004 (2015).

\bibitem{Wineland1998} D. J. Wineland, C. Monroe, W. M. Itano, D. Leibfried, B. E. King, and D. M. Meekhof, J. Res. Natl. Inst. Stand. Technol. \textbf{103}, 259 (1998).

\bibitem{Hafner2008} H. H\"afner, C. F. Roos, and R. Blatt, Phys. Rep. \textbf{469}, 155 (2008).

\bibitem{James1998} D. F. V. James, Appl. Phys. B \textbf{66}, 181 (1998).

\bibitem{Marquet2003} C. Marquet, F. Schmidt-Kaler, and D. F. V. James, Appl. Phys. B \textbf{76}, 199 (2003).

\bibitem{Campos1989} R. A. Campos, B. E. A. Saleh, and M. C. Teich, Phys. Rev. A \textbf{40}, 1371 (1989).

\bibitem{Caves1991} C. M. Caves, C. Zhu, G. J. Milburn, and W. Schleich, Phys. Rev. A \textbf{43}, 3854 (1991).

\bibitem{yi1996} F. Hong-yi and F. Yue, Phys. Rev. A \textbf{54}, 958 (1996).

\bibitem{Lee2005} P. J. Lee, K.-A. Brickman, L. Deslauriers, P. C. Haljan, L.-M. Duan, and C. Monroe, J. Opt. B \textbf{7}, 371 (2005).

\bibitem{Schneider2012} C. Schneider, D. Porras, and T. Schaetz, Rep. Prog. Phys. \textbf{75}, 024401 (2012).

\bibitem{Ivanov2016} P. A. Ivanov, N. V. Vitanov, and K. Singer, Sci. Rep. \textbf{6}, 28078 (2016).

\bibitem{Ivanov2016_1} P. A. Ivanov, Phys. Rev. A \textbf{94}, 022330 (2016).

\bibitem{Lipkin1965} H. J. Lipkin, N. Meshkov, and A. Glick, Naucl. Phys. \textbf{62}, 188 (1965).

\bibitem{Wineland1979} D. J. Wineland and W. M. Itano, Phys. Rev. A \textbf{20}, 1521 (1979).

\bibitem{Braunstein1994} S. L. Braunstein and C. M. Caves, Phys. Rev. Lett. \textbf{72}, 3439 (1994).

\bibitem{Paris2009} M. G. A. Paris, Int. J. Quantum. Inf. \textbf{7}, 125 (2009).

\bibitem{Pezze2014} L. Pezze and A. Smerzi, "Quantum theory of phase estimation", in G. M. Tino and M. A. Kasevich (Eds.), Atomic Interferometry. Proceedings of International School of Physics Enrico Fermi, Course 188, Varenna, 691-741, IOS Press (2014).


\end{thebibliography}
\end{document}